\documentclass[aps,pra,twocolumn,
groupedaddress,superscriptaddress,
bibnotes,amsfonts,longbibliography,
citeautoscript,a4paper,
floatfix]{revtex4-2}
\usepackage{latexsym}
\usepackage{amsmath}
\usepackage{amssymb}
\usepackage{graphicx}
\usepackage[T1]{fontenc}
\usepackage[open]{bookmark}
\usepackage{orcidlink}
\usepackage{hyperref}
\hypersetup{colorlinks=true,allcolors=blue}
\usepackage{xcolor} 
\usepackage{newtxtext}
\usepackage{dcolumn}% Align table columns on decimal point
\usepackage{bm}% bold math
\begin{document}

\preprint{APS/123-QED}

\title{Gibbs Phenomenon and Friedel Oscilations: Similarities, Differences, and the Educational Potential of their Comparison}% Force line breaks with \\

\author{Christos Mystilidis\,\orcidlink{0000-0002-8011-2154}}
\email{chrmys@mci.sdu.dk}
\affiliation{POLIMA---Center for Polariton-driven Light-–Matter Interactions, University of Southern Denmark, Campusvej 55, DK-5230 Odense M, Denmark}
\affiliation{School of Electrical and Computer Engineering, National Technical University of Athens, 9 Iroon Polytechniou Street GR 157-73 Zografou, Athens, Greece}
\author{Christos Tserkezis\,\orcidlink{0000-0002-2075-9036}}
\affiliation{POLIMA---Center for Polariton-driven Light-–Matter Interactions, University of Southern Denmark, Campusvej 55, DK-5230 Odense M, Denmark}
\affiliation{D-IAS---Danish Institute for Advanced Study,\\
University of Southern Denmark, Campusvej 55, DK-5230 Odense M, Denmark}
\author{George Fikioris\,\orcidlink{0000-0002-0987-8835}}
\affiliation{School of Electrical and Computer Engineering, National Technical University of Athens, 9 Iroon Polytechniou Street GR 157-73 Zografou, Athens, Greece}%

\date{\today}% It is always \today, today,

\begin{abstract}
We explore the similarities and differences between the Gibbs phenomenon in partial Fourier representations of discontinuous signals and Friedel oscillations in the electron density of a solid near an anomaly. Inspired by the apparent similarities of the two phenomena, we perform a detailed exploration of both from the viewpoint of an engineer being introduced to a concept from solid-state physics. Focusing on the density of an one-dimensional electronic gas confined by a square potential, we show that, despite the similarities, Friedel oscillations cannot be attributed to the inability of a partial Fourier series to describe a discontinuity. Nevertheless, the two phenomena do exhibit similarities, which can be exploited to develop intuition. By adopting an educational style, we hope to establish some common language between electrical engineers and condensed-matter physicists, hoping that this can further inspire the two communities to seek intuition and further comprehension within neighbouring but different disciplines.

\end{abstract}

\keywords{Gibbs Phenomenon, Friedel Oscillations}
%Use showkeys class option if keyword
\maketitle

%\tableofcontents

\section{\label{sec:intro} Introduction}

The Gibbs phenomenon (also called the Gibbs--Wilbraham phenomenon~\cite{Hewitt1979}) is well-known to engineers and applied mathematicians. Undergraduate students, doctoral-level researchers or frontline production engineers routinely encounter it, in signal- or Fourier-analysis courses~\cite{MIT2024,EPFL2025,Warwick2025,Delft2024}, as a research subject on its own~\cite{Fikioris2017} or as an engineering challenge which can give rise to artifacts in biomedical images~\cite{Peronne2015,Zhao2020,Wang20223}  or to distortions in digital signal processing~\cite{Pan2001,Lin2019,Liu2022}. The Gibbs phenomenon also comes up in diffraction~\cite{Milla2021,Shimobaba2025}, Computational Electromagnetics~\cite{Perez2007}, antenna engineering~\cite{Cho2025}, and (more related to the context of this work) to numerical approximations of the step-like Fermi--Dirac distribution in time-dependent quantum dynamics calculations~\cite{Lothman2021}. The Gibbs phenomenon is often a nuisance that needs to be suppressed, or at least to be properly taken into account.

Friedel oscillations, predicted in the 50s by Friedel~\cite{Friedel1952} and first observed four decades later by Crommie and 
coworkers~\cite{Crommie1993} and independently by Hasegawa and Avouris~\cite{Hasegawa1993},
are a standard concept of solid state physics, but perhaps not routinely encountered in most engineering curricula. They describe the response of the charge density of a fermionic system to an impurity, and are typically visualised through Scanning Tunneling Microscopy (STM). STM snapshots of the local density of states of a fermionic gas in the proximity of an impurity do justice to the phrase ``electron sea'', revealing intricate landscapes of density standing waves~\cite{Hasegawa1993,Crommie1993}. Friedel oscillations play a central role in mapping complicated Fermi surfaces (through Fourier-transforming images of STM)~\cite{Sprunger1997,Petersen1998}---a mapping that, aside from its superlative importance in understanding the electronic properties of materials, is also a highly aesthetic endeavour~\cite{Dugdale2016}. Such oscillations have been detected and analysed in a wealth of systems including graphene-based architectures~\cite{Cheianov2006,Yin2023}, transition metal dichalcogenides (TMDs)~\cite{Siegl2025}, and topological insulators~\cite{Stephanovic2022}, aside from metals and doped semiconductors~\cite{Crommie1993,Hasegawa1993,Wielen1996}, which are the archetypical instances of materials that support them.

\begin{figure*}[t!]
    \centering
    \includegraphics[width=\linewidth]{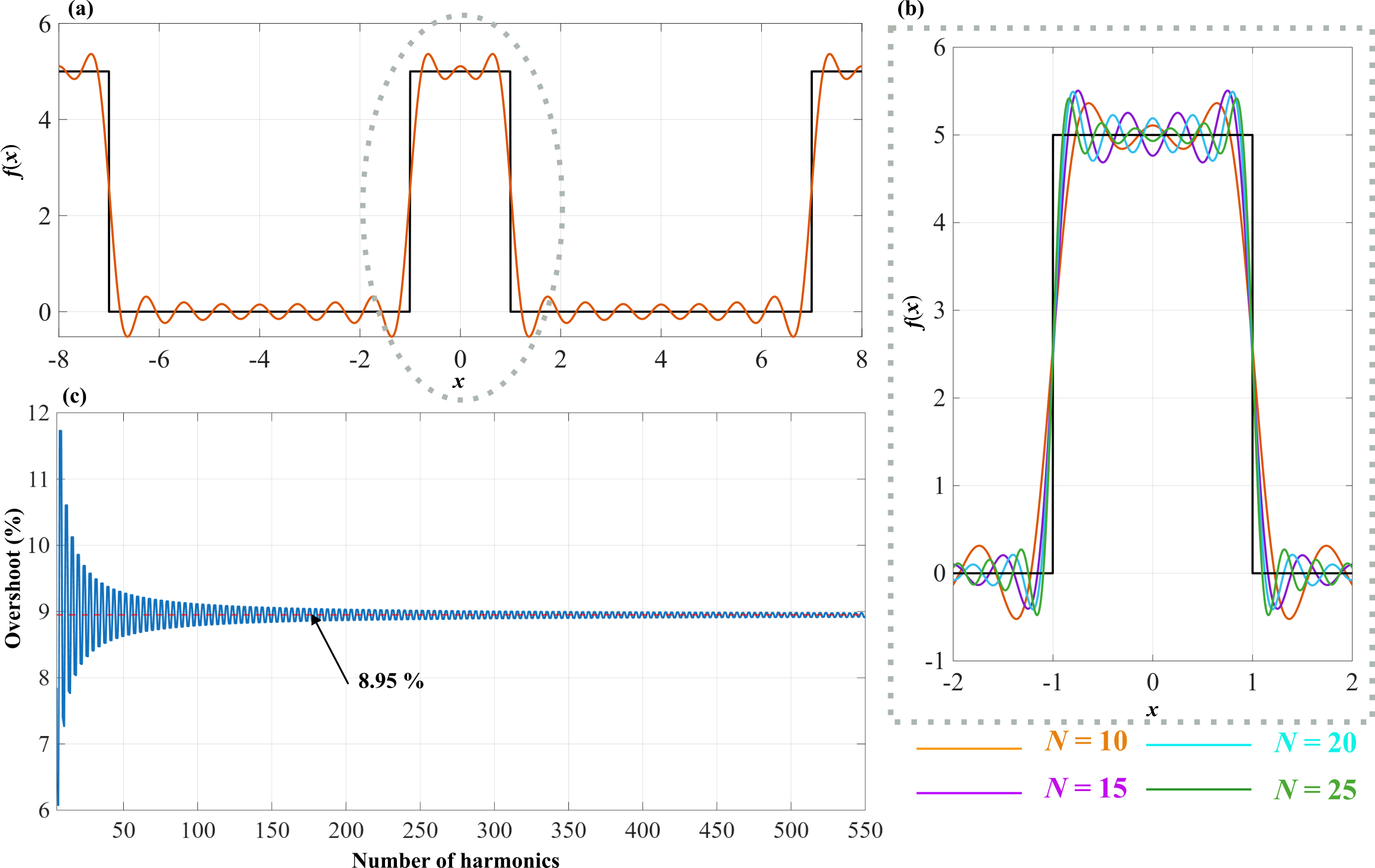}
    \caption{Demonstration of the Gibbs phenomenon. (a) Representation of a square wave $f(x)$ (black line) by its partial Fourier series (orange line). The square has an amplitude of $5$, and $10$ harmonics are used to represent it. The oscillatory behavior is evident throughout. (b) Zoom-in around the central pulse of the square wave. We progressively increase the number $N$ of harmonics used to reconstruct the pulse, but the Gibbs phenomenon persists. (c) Amplitude of the most pronounced overshoot with increasing number of terms (from $5$ to $550$). The amplitude converges to the well-known $8.95\%$ value.}
    \label{fig:square}
\end{figure*}

Despite residing in distinct territories of science, the two phenomena share striking visual and essential similarities, echoing sharp truncations of the Fourier space. A number of works~\cite{Zhang2001,Politano2013,Samoilenka2020,Talchakov2023} has already recognised similarities but without elaborating further. As such, the goal of this article it to unravel the relation between the Gibbs phenomenon and Friedel oscillations. Starting from the obvious similarities, we proceed to show that the two phenomena are not the same, that is, Friedel oscillations are not physical manifestations of the mathematical Gibbs phenomenon. Yet, their similarities can give rise to potent educational tools.  For the engineer or the applied mathematician who has digested the lessons of the Gibbs phenomenon, behaviours of Friedel oscillations, such as the influence of dimension or temperature, can be understood in a straightforward manner. 
The paper adopts an educational approach, hoping to establish a common language for the engineering and condensed-matter disciplines, and inspire mutual adoption of each other's tools and intuitions.
\begin{figure*}[t!]
    \centering
    \includegraphics[width=1.0\linewidth]{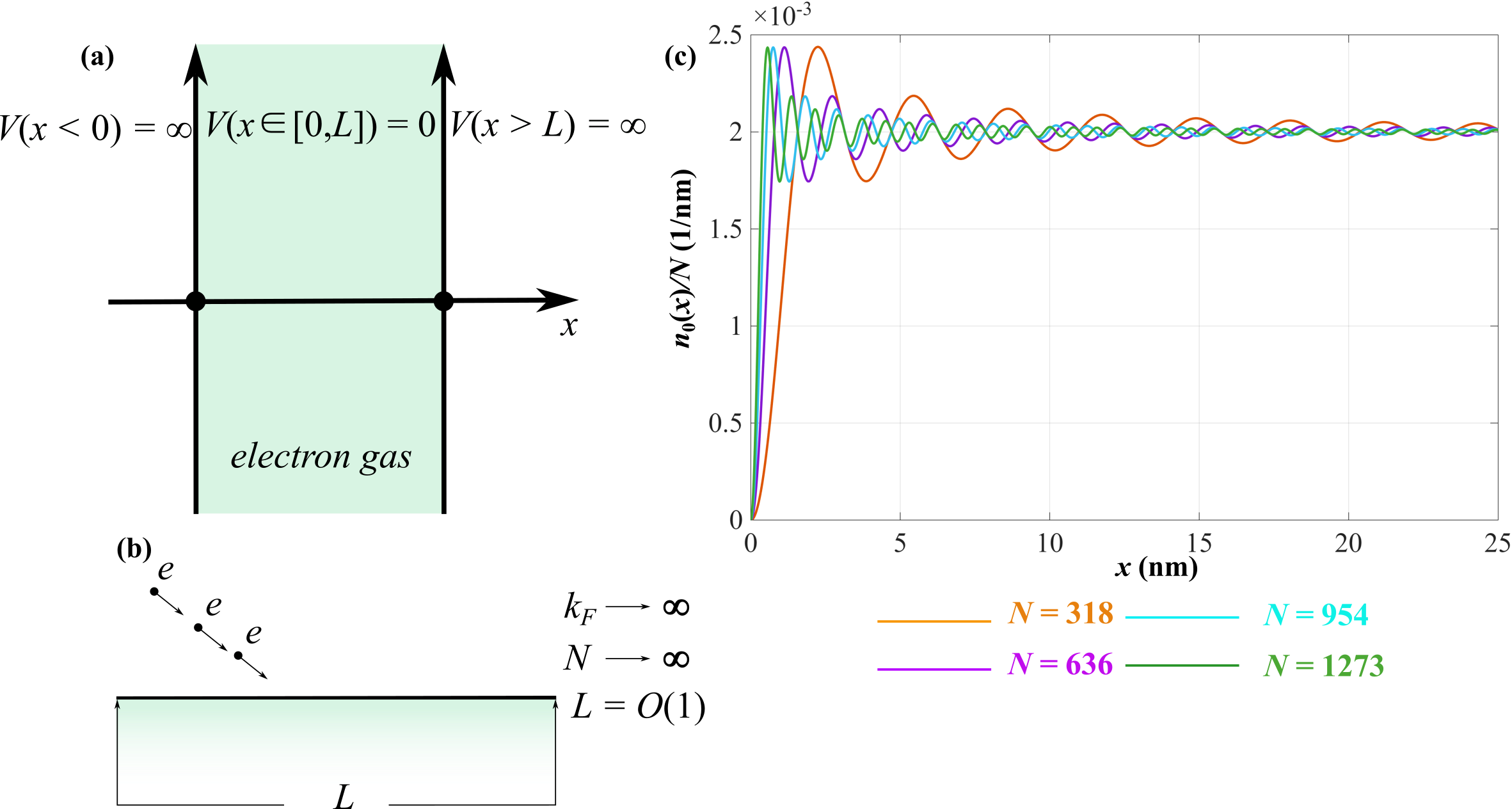}
    \caption{Demonstration of Friedel oscillations. (a) Toy problem that occupies most of our study: An 1D electron gas is confined in $[0,L]$ due to the action of an infinite potential outside of this interval. (b) Physical demonstration of how the number of states is increasing. The length of the electron gas is kept fixed and the Fermi wavenumber increases, which is equivalent to pumping electrons in a finite system. (c) Friedel oscillations for the system (a) apparent in the normalised (to the number of states) ground-state density over the the distance inside the electron gas. The phenomenology is very similar to that of the Gibbs phenomenon in Figure \ref{fig:square}(b).}
    \label{fig:friedel}
\end{figure*}
\section{\label{sec:fund} Fundamentals}
\subsection{\label{subsec:gibbs}The Gibbs phenomenon}
The variety of research disciplines that engage with the Gibbs phenomenon originates from the omnipresence of Fourier analysis in engineering and mathematics, and can be understood by providing a clear description of it~\footnote{Although in this work we echo the most common incarnation of the Gibbs phenomenon, that arising from Fourier series, we clarify that it can appear also in multiple different expansions, see for example~\cite{Gottlieb1997}.}. Thus \emph{the Gibbs phenomenon is the inability of a partial Fourier series to accurately reconstruct a piecewise smooth function, irrespective of the number of terms added, near the point of irregularity and manifesting as an oscillatory behaviour}. The oscillations are most evident close to the irregular point (most often a jump discontinuity, but also a logarithmic singularity~\cite{Boyd2009}; in this work we are concerned with the former case), and the partial Fourier series overshoots/undershoots the value of the ``final function'', i.e. of the function it intends to reproduce. It is crucial to underline that the Gibbs phenomenon is completely resolved, should we take the full series (that is, should we add infinite terms)~\cite{Fikioris2017}. The involved Fourier series then converges uniformly to the correct function in all closed intervals not containing discontinuity points and, at discontinuity points, pointwise to the midpoint of the jump~\cite{Jerri1998}. There is a \emph{final function} that can theoretically be retrieved in this manner. On the other hand, as long as a finite but increasing number of terms is added, we observe an increase of the frequency of the oscillatory behaviour and the squeezing thereof close to the point of discontinuity. In this large-$N$ limit (where $N$ is the number of terms in the Fourier series), the Gibbs phenomenon acquires the often quoted value of the first overshoot, namely around $8.95 \%$ of the size of the jump of discontinuity~\cite{Jerri1998}, an overshoot that cannot be resolved by further increasing $N$.

In Fig.~\ref{fig:square}(a), we depict perhaps the most standard case of the Gibbs phenomenon, arising when a square wave, here described by a function $f(x)$, is expanded in cosine Fourier series \cite{Hewitt1979,Fikioris2017,Jerri1998}. The oscillatory behavior is clear. Increasing $N$ leads to a better description far from the jump of discontinuity; the overshoot however \emph{never} vanishes, as is clear in Fig.~\ref{fig:square}(b). After a sufficient amount of terms are included in the series, the amplitude of the overshoot converges to $8.95\%$ of the jump. This can be seen in Fig.~\ref{fig:square}(c). In~\cite{Fikioris2017} it was shown that such value of the overshoot is just the zero order term of expansions that depend on powers of $1/n$; these manifest in Fig.~\ref{fig:square}(c)
~\footnote{Though notice that in any numerical implementation the phenomenon may be somewhat blurred by other parameters, notably the sampling of real space $x$ or the order and the type of numerical integration needed to calculate the Fourier coefficients (even if it is unnecessary in this example).}. This is the essence of the statement \emph{the Gibbs phenomenon is a large-$N$ effect}. For the history of the Gibbs phenomenon,  see the delightful work by Hewitt and Hewitt~\cite{Hewitt1979} and also the more recent book by Jerri~\cite{Jerri1998}. The Gibbs phenomenon is a mathematical phenomenon and constitutes the first pillar of this work.

\subsection{\label{subsec:friedel}Friedel oscillations}
The other pillar of this work, Friedel oscillations, occurs in fermionic systems at the presence of some sort of disorder (which can be a lattice vacancy, an impurity, or simply the breaking of the translational invariance of the bulk due to an interface) and manifest notably in the ground-state (henceforth electron) density of such systems. Friedel oscillations are noticeable near the disorder, and decay algebraically away from it, the strength of such decay being determined by multiple factors such as the dimensionality of the system at hand~\cite{Siklitskaya2025}. Their wavelength is closely related to the Fermi wavelength $\lambda_F$ (and wavenumber $k_F$), which for an isotropic free-electron gas equals half its value $\lambda_F/2 = \pi/k_F$~\cite{Petersen1998}. Friedel oscillations can arise very simply for toy problems (and we will consider such later), however their general manifestation is intuitive. In the bulk, the electron gas density is a superposition of plane waves, those corresponding to the wavefunctions of all electrons in the gas. In the presence of disorder at a point or region of space, the electron waves get reflected, interfere with themselves, and create a standing-wave pattern~\cite{Dobson1995,Sessi2015,Bena2016,Shen2024,Siklitskaya2025} (see Fig.~\ref{fig:friedel}(c) for the simple system of Fig.~\ref{fig:friedel}(a)). More rigorously, Friedel oscillations are discussed in the context of electron screening at the level of Lindhard theory~\cite{Ashcroft1976}. Unlike Thomas--Fermi theory, which is pertinent for sufficiently low transferred momenta, and where the electron gas screens efficiently disturbances of the external potential, Lindhard screening reveals both more interesting structuring and the existence of long-range interactions~\cite{Ashcroft1976,Sprunger1997,Cheianov2006,Bouhassoune2014}. The reason behind Fridel oscillations is then singularities in the Lindhard function in the reciprocal space, which describes the electronic response to external perturbations, and which translate to oscillatory behaviour in the real-space electron density through Fourier transform~\cite{Ashcroft1976,Sprunger1997,Simion2005}.

Friedel oscillations are closely intertwined with the geometry of the Fermi surface. In fact, another explanation behind their generation (and the reason behind the nonanalyticity of the Lindhard function) stems from the \emph{sharpness} of the Fermi surface (at absolute zero)~\cite{Dobson1995,Gabovich1978,Liebsch1997}. This is understood, since the Fourier transform of the electron density can be used to map the Fermi surface~\cite{Petersen1998}. The shape thereof bequeaths its characteristics to Friedel oscillations (which can then demonstrate  anisotropic, highly directional patterns~\cite{Hosur2012,Lawlor2013,Bouhassoune2014,Farajollahpour2018}). We find that the argument regarding the abrupt termination of the Fermi surface is what invites the comparison between Friedel oscillations and the Gibbs phenomenon: the concept of a truncated and sharp Fourier space permeates both phenomena and lies at the heart of their oscillatory nature. It is the reason behind the obvious similarities we notice in Fig.~\ref{fig:friedel}. In particular, in Fig.~\ref{fig:friedel}(c) we observe, for the simple one-dimensional (1D) problem depicted in Fig.~\ref{fig:friedel}(a), that as we increase the number of states (the Fermi wavenumber, see also Fig.~\ref{fig:friedel}(b)), the oscillations become faster and get squeezed close to the boundary. Note that the ground-state density grows with the number of states, and as such we present above the density normalised to this number (the normalised and un-normalised densities exhibit an essential difference as we show in the next Section). The oscillations do not seem to vanish in this limited example. Our discussion below Fig.~\ref{fig:square} about the Gibbs phenomenon is very similar, and we cannot but seriously consider a deep connection or even overlap between the two phenomena.

\section{\label{sec:gibbsfriedel} Gibbs Phenomenon is not Friedel Oscillations (and vice-versa)}
\subsection{\label{subsec:demo_1d}A simple problem: perfectly confined 1D electron gas }
Our platform for evaluating whether Friedel oscillations are physical manifestations of the Gibbs phenomenon will be the 1D non-interacting electron gas shown in Fig.~\ref{fig:friedel}. In this work, we assume that $L \gg \lambda_F/2$ (and similarly for all lengths along the confined direction) in order to simulate a semi-infinite gas that is abruptly terminated by an interface. We use a very simple potential to confine the electron gas in $ x \in [0,L]$, illustrated in Fig.~\ref{fig:friedel} as well, which induces a Dirichlet boundary condition, namely $\psi(0)=\psi(L)=0$. The electronic wavefunctions are reflected on the hard walls, interfere with their incident components, thus creating Friedel oscillations. Textbook derivations allow for analytic determination of the wavefunctions of the system~\cite{Griffiths2018},
\begin{equation}
    \label{eq:notgibbs_wavn}
    \psi_n (x) = \sqrt{\frac{2}{L}} \sin{\left( \frac{\pi n}{L} x\right)},
\end{equation}
where $n$ is a positive integer and $k_n = \pi n/L$ are the wavenumbers of each discrete state. By virtue of this proportionality, we can interchange $n$ and $k$ as indices of the wavefunctions (and later of the summations). Notice that beyond the infinite wall, at $x \notin [0,L]$, the wavefunction $\psi_n(x) =0$.

We construct the ground-state density by taking~\cite{Liebsch1997}
\begin{equation}
    \label{eq:notgibbs_density}
    n_0(x) = 2 \sum_k f_k \left|\psi_k (x) \right|^2,
\end{equation}
where the factor $2$ is the electron degeneracy factor, reflecting the Pauli exclusion principle, and
the summation goes over all acceptable wavenumbers (states), in turn defined by $f_{k}$, which represent the Fermi--Dirac distribution. In this section we consider absolute zero temperature, $T=0$\,K, and as such $f_k=\Theta(k_F - k)$~\cite{Ashcroft1976}, where $\Theta(x)$ is the Heaviside function. 

The Fermi--Dirac distribution allows us to reflect: a function with jump discontinuity has appeared in our formalism. Yet, it does not correspond to the ``reconstructed'' function (density). It is, however, critical, because it constitutes the reason behind the finite and sharp Fermi sea, that is, of a sharp Fourier space. The situation is very similar to that of the conditions behind the Gibbs phenomenon and invites confusion; however, we must clarify that in the Gibbs phenomenon the discontinuity exists in real space, \emph{and} the Fourier space is sharply truncated. 

We return to Eq.~(\ref{eq:notgibbs_density}), and use Eq.~(\ref{eq:notgibbs_wavn}). After some simple algebra, the result becomes~\footnote{The density is defined likewise in $[0,L]$ rather than $x \geq 0$. However, since we assumed a sufficiently long electron gas, we treat it as semi-infinite, eventually focusing on the perturbation that the confinement at $x=0$ causes. Notice further that the density displays even symmetry with respect to $x=L/2$, hence all our discussions can readily be mirrored for the oscillations close to $x=L$.}
\begin{equation}
    \label{eq:notgibbs_dir}
    n_0(x) = \frac{2N}{L} - \frac{2}{L} \underbrace{\sum_{n=1}^{N} \cos{\left( \frac{2 \pi n}{L} x \right )}}_{D_N(x)/2-1/2}, \quad x \geq 0,
\end{equation}
where $N = \lfloor k_F L/\pi \rfloor$ is the maximum number of states for the 1D electron gas and $\lfloor...\rfloor$ stands for the floor operator. The first term ($2N/L$) corresponds to the bulk contribution to the ground-state density, while the second reflects the Friedel part. We make two points here: (a) Eq. (\ref{eq:notgibbs_dir}) corresponds to a peculiar Fourier cosine series. If the reconstructed ground-state density demonstrates a jump of discontinuity (in real space), then this expression could give rise to the Gibbs phenomenon. (b) A core element of Gibbs phenomena (and of partial Fourier series in general) appears in Eq.~(\ref{eq:notgibbs_dir}). This is the Dirichlet kernel $D_N(x)$~\cite{Jerri1998}, a highly oscillatory function with a pronounced peak at $x=0$; it is instrumental in extracting the positions of local undershoots/overshoots that appear in the Gibbs phenomenon, as well as the amplitudes thereof. Smoothing this kernel (e.g., by some sort of averaging, see the Fej\'{e}r kernel~\cite{Fikioris2017,Gottlieb1997,Jerri1998}) is a standard procedure to alleviate the Gibbs phenomenon when numerically evaluating truncated Fourier series.

On the other hand, assuming that the number of states~\footnote{The asymptotics here must be performed with caution: in particular, we increase $N$ by increasing $k_F$, while keeping the length $L$ of the electron gas fixed. The physical situation is shown in Fig.~\ref{fig:friedel} and corresponds to pumping electrons on a finite system. This is not identical to the case where the length increases but $k_F$ remains fixed. Our line of reasoning is that we aim for a summation over infinite discrete states instead of a continuum of states.} $N \rightarrow \infty$, we notice that the first term explodes, which is intuitive, as a system with infinite states and fixed size should have infinite density. The second term diverges for $x=0$ like $2N/L$ and hence cancels with the first term (reflecting the boundary conditions), while for $x > 0$ it is bounded and hence the whole density diverges. There is no final function that we can recover in this manner. Each individual ground-state density depends on $N$, and cannot be thought as the partial Fourier series of a final function (and this is physically anticipated, as increasing or in general changing $N$ corresponds to a different physical system) that has any spatial discontinuity. The oscillatory pattern does not arise from the inability of the cosine Fourier series above to describe a final function near a discontinuity but, importantly, from the abrupt truncation of the Fermi sea. Friedel oscillations have a similar mathematical background as the Gibbs phenomenon, however the argument above, for an example that favours the comparison with the Gibbs phenomenon, is strong evidence, in our view, against the two phenomena being identical.

\subsection{\label{subsec:norm}Normalised ground-state density}
One might insist on normalising the density. For a fixed $L$, the resulting function should be bounded and not diverge like the physical density. Using further the closed-form formula for the Dirichlet kernel leads to 
\begin{equation}
    \label{eq:notgibbs_norm_closed}
    \frac{n_0(x)}{N} = \frac{2}{L} - \frac{1}{NL} \left[ \frac{\sin{\left( (2N+1) \frac{\pi x}{L} \right)}}{\sin{\left( \frac{\pi x}{L} \right)}} -1  \right], \quad x \geq 0.
\end{equation}
Taking then $N \to \infty$ yields 
\begin{equation}
    \label{eq:notgibbs_asym1}
    \frac{n_0(x)}{N} \sim 
    \begin{cases}
       \frac{2}{L}, \quad x > 0, \\
       0, \hspace{0.43cm} x=0.
    \end{cases}
\end{equation}
Remembering additionally that $n_0(x) = 0$ for $x<0$, leads to 
\begin{equation}
    \label{eq:notgibbs_asym2}
    \frac{n_0(x)}{N} \sim \frac{2}{L} \Theta(x), \quad (N \to \infty).
\end{equation}
The reconstructed density (which now exists) has a very notable discontinuity (at $x=0$) in real space. The closed-from expressions can be easily manipulated to study the convergence of the finite $N$ expressions to the limiting value. Consider the remainder
\begin{equation}
    \label{eq:notgibbs_lim}
    r_N(x) = \left[\frac{n_0(x)}{N}\right]^{(\infty)} - \left[\frac{n_0(x)}{N}\right]^{(N)}.
\end{equation}
As $N \to \infty$  the limit of this remainder is zero for any \emph{nonzero} $x$, implying pointwise convergence away from the discontinuity. In fact, $r_N(x) = O(1/N)$ for $x > 0$, so the aforementioned convergence is slow.  In addition, we will show
that the quantity $\max_{x>0} r_N(x)$ does not have limit zero, signalling an overshoot that resembles that of the Gibbs phenomenon. This is as close to Gibbs as we can get in the present analysis. What compromises it is that the new ``Fourier'' coefficients depend on the very number of terms added $N$ (and hence are not Fourier coefficients). In what follows, we calculate this quantity explicitly and find the said overshoot.  We clarify that the overshoot calculated is particularly dependent on the details of the problem (e.g., the shape of the confining potential and as later seen, dimensionality and temperature) and we do not expect our results to hold universally (unlike the Gibbs overshoot).

\begin{figure*}[t!]
    \centering
    \includegraphics[width=1.0\linewidth]{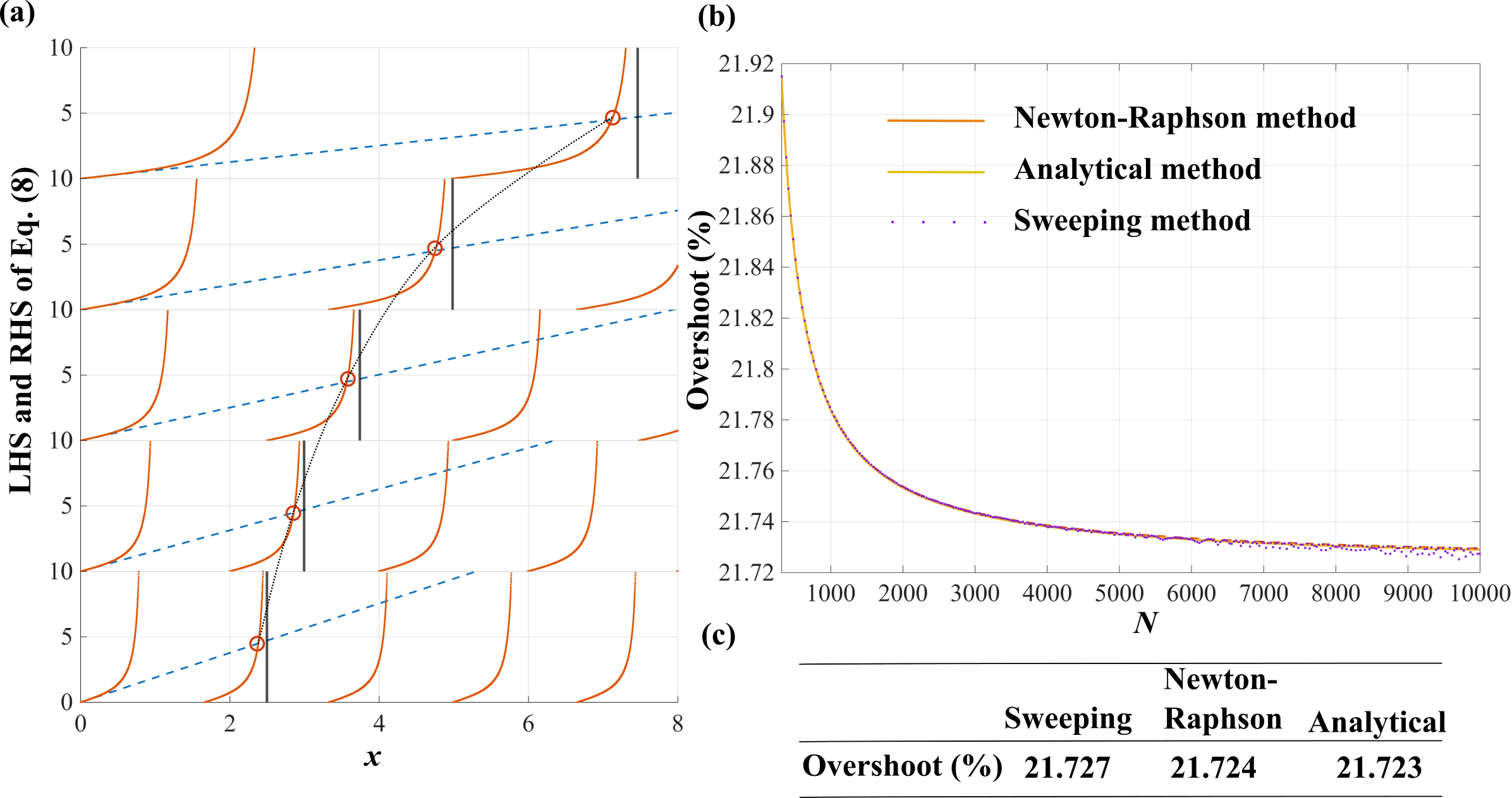}    
    \caption{(a) Graphical depiction of the line of reasoning of the solution of Eq.~(\ref{eq:notgibbs_deriv}). With orange lines we depict the RHS of Eq.~(\ref{eq:notgibbs_deriv}), while with dashed blue the LHS. In the five snapshots shown, we progressively increase $N=100,150,200,250,300$. The vertical asymptote $x=3L/[2(2N+1)]$ is denoted by a black line. Finally, the sought after solution is denoted by red circles and its evolution with $N$ with a dotted black curve, which is a guide to the eye. It is clear that the solution is pushed towards zero, following the vertical asymptote. (b) Evolution of the first overshoot of $n_0(x)/N$ using three different methods. In orange, the result comes from the following method: the solution of Eq.~(\ref{eq:notgibbs_change}) is determined through a Newton--Raphson method (starting point $z=4.3$), then it is translated to $x$ space, which is then substituted in Eq.~(\ref{eq:notgibbs_norm_closed}). In yellow, the overshoot as derived from Eq.~(\ref{eq:notgibbs_asumpsol_pos}) and Eq.~(\ref{eq:notgibbs_norm_closed}) is plotted. Finally, in purple dots, the overshoot is calculated by sweeping Eq.~(\ref{eq:notgibbs_density}) in $[0,L]$, discretised in $1500000$ points, collecting the first maximum. (c) Numerical value of the overshoot for $N=9994$. The agreement between the different methods is excellent. The spread of the sweeping solution as $N$ increases, owing to the progressive inability of the dense but fixed discretisation of $[0,L]$ to resolve in a satisfactory manner the decreasing wavelength of Friedel oscillations.}
    \label{fig:overshoot}
\end{figure*}

We first calculate the first derivative of $n_0(x)$ as given by Eq.~(\ref{eq:notgibbs_norm_closed}), and require it to be zero in order to locate extrema (and hence overshoots), leading to~\footnote{For the derivation of Eq.~(\ref{eq:notgibbs_deriv}) we exclude several points; we do not investigate here whether these correspond to solutions of Eq.~(\ref{eq:notgibbs_deriv}), as we underline that we only seek the first nonzero one, that corresponds to the maximum overshoot. The excluded points are: $x \neq \rho L,\rho L + L/2,(\rho L + L/2)/(2N+1) $, with $\rho$ a positive integer.}
\begin{equation}
    \label{eq:notgibbs_deriv}
    (2N + 1) \tan{ \left( \frac{\pi x}{L} \right) } = \tan{ \left[  (2N + 1) \frac{\pi x}{L}   \right]}.
\end{equation}
This is a transcendental equation and not amenable to analytical progress. Using a simple line of reasoning and asymptotics we can, however, bracket the solution and then determine an approximation of it. The left-hand side (LHS) tangent has a period of $L$ and is positive in $(0,L/2)$. The right-hand side (RHS) tangent varies much faster with a period of $L/(2N+1)$. As $N \to \infty$, the period becomes vanishingly small. Since the tangent function changes sign over one period and the RHS tangent evolves faster than the LHS one, we expect to find the first nonzero solution (corresponding to the maximal overshoot) in the first few periods of the RHS tangent. The periods (and by extension the sought solution) get squeezed close to zero as $N \to \infty$ implying $x \ll L/\pi$. We show graphically our rationale in Fig.~\ref{fig:overshoot}(a) and Taylor expand the LHS tangent leading to 
\begin{equation}
    \label{eq:notgibbs_taylor}
    (2N + 1) \frac{\pi x}{L} = \tan{ \left[  (2N + 1) \frac{\pi x}{L}   \right]},
\end{equation}
or, by setting $z=(2N + 1) \pi x/L$
\begin{equation}
    \label{eq:notgibbs_change}
    z = \tan{ z }.
\end{equation}
For the first period of the RHS, it is simple to prove that $z < \tan{z}$. Using the Bolzano theorem for $z \in [\pi,3\pi/2)$---or $x \in [L/(2N+1),3L/[2(2N+1)])$---it is simple to show that there exists at least one solution in this interval and, since $f(z)=z-\tan{z}$ is strictly monotonous there, the solution is further unique. We would like then to show that, as $N \to \infty$, the solution tends to the vertical asymptote of the bracketing interval, which in $z$ space is $3 \pi/2$. To do so, we assume a solution that is $z_s = 3 \pi/2 - \varepsilon_z $, where $\varepsilon_z>0$ is a small positive number. Substituting $z_s$ into Eq.~(\ref{eq:notgibbs_change}) leads to
\begin{equation}
    \label{eq:notgibbs_asumpsol}
    \frac{3\pi}{2} - \varepsilon_z -\tan{\left(  \frac{3\pi }{2} - \varepsilon_z  \right)} = 0,
\end{equation}
which, upon Laurent-expanding the tangent, becomes
\begin{equation}
    \label{eq:notgibbs_epsilon}
    \frac{3\pi }{2} - \frac{1}{\varepsilon_z} = 0.
\end{equation}
or~\footnote{It is simple to show that $\varepsilon_x \to 0$ as $N \to \infty$ and the solution approaches the asymptote.}
\begin{equation}
    \label{eq:notgibbs_asumpsol2}
    z_s = \frac{9 \pi^2 - 4}{6 \pi}.
\end{equation}
In $x$ space, we have found an approximation for the solution, namely
\begin{equation}
    \label{eq:notgibbs_asumpsol_pos}
    x_s = \frac{L}{(2N + 1) \pi}\frac{9\pi^2-4}{6\pi}.
\end{equation}
Using next $x_s$ in Eq.~(\ref{eq:notgibbs_norm_closed}), and after some algebra, we find that
\begin{equation}
    \label{eq:notgibbs_norm_xs}
    \frac{n_0(x_s)}{N} \sim \frac{2}{L} - \frac{2}{L} \frac{\sin{\left( \frac{9 \pi^2 - 4}{6 \pi} \right)}}{\frac{9 \pi^2 - 4}{6 \pi}}.
\end{equation}
The overshoot is
\begin{equation}
    \label{eq:notgibbs_overshoot}
    \text{Overshoot}(x_s) = \left|  \frac{r_N(x_s)}{\left[\frac{n_0(x_s)}{N}\right]^{(\infty)}}   \right| \approx 21.72 \%.
\end{equation}
We have confirmed the validity of this result using (a) a Newton--Raphson method to numerically solve Eq.~(\ref{eq:notgibbs_change}) and (b) a na\"{i}ve sweeping method, to locate the total maximum in Eq.~(\ref{eq:notgibbs_norm_closed}). We summarise the results in Fig.~\ref{fig:overshoot}. 

Although the overshoot is quite larger than that of the Gibbs phenomenon, this is not our primary objection to the equivalence between Friedel oscillations and the Gibbs phenomenon. The inability to find a limiting function, and as such the lack of meaning of a limiting procedure (let alone of a discussion on convergence), are all we need to discard the equivalence. However, the two phenomena are indeed very close mathematically, notably so in the normalised case. We exploit this relation next to extract intuitive explanations about the behaviour of Friedel oscillations in higher-dimensional systems and in 1D systems at nonzero temperature.

\section{Using the similarities to build intuition}

In this Section, we discuss in a colloquial and mathematically relaxed manner, how the significant parallels between the two phenomena can be exploited to offer intuitive explanations to understand physical behaviors of Friedel oscillations when it comes to the effect's dependence on dimensions and temperature. What allows such discussions is, of course, the Fourier analysis of spectrally sharp features.

\subsection{Influence of dimensions}

As a first step, we study a perfectly confined electron gas in higher dimensions. It is well-known that the amplitude of Friedel oscillations depends on, and is weakened, as the dimensions increase~\cite{Siklitskaya2025,Simion2005,Gabovich1978} (their long-range decay being dictated by the $1/r^d$ rule, where $d$ represents the dimensions of the electron gas). From a physical viewpoint, higher dimensionality enlarges the set of allowed momentum directions, thus generating oscillations with different (less than $2k_F$) wavenumbers, which then interfere destructively and suppress the dominant $2k_F$ contribution, leading to faster decay.

We start with a two-dimensional (2D) electron gas, which is  perfectly confined in $[0,L_x]$ like its 1D counterpart and extends infinitely along $y$ (simulated by periodic boundary conditions, with imposed period $L_y$), a standard textbook treatment of the system~\cite{Bruus2016}. Separation of variables allows for analytical determination of the wavefunctions and hence of the electron                                   density
\begin{equation}
    \label{eq:pedagdim_2D}
    n_0(x,y) = \frac{2}{L_x L_y} \sum_{n_x} \sum_{n_y}1 - \frac{2}{L_x L_y}\sum_{n_x} \sum_{n_y} \cos{\left( \frac{2 \pi n_x}{L_x} x \right )},
\end{equation}
where $n_x$ is a natural number and $n_y$ an integer one.  The wavenumbers are confined within the Fermi circle \mbox{$k_x^2 + k_y^2 \leq k_F^2$} and are discretised due to the aforesaid boundary conditions, namely $k_x=\pi n_x/L_x$ and $k_y=2\pi n_y/L_y$. The Fermi sea in this case is a disk. By this token, it is possible to rewrite Eq.~(\ref{eq:pedagdim_2D}) in terms of a single sum,
\begin{equation}
    \label{eq:pedagdim_2D_1sum}
    \begin{split}
        & n_0(x,y) =  n_0^{\rm(2D)} -  \\ 
                   & \frac{2}{ L_x L_y}\sum_{n_x=1}^{N_{x}} \left( 2 \underbrace{\left \lfloor \frac{L_y}{2\pi}\sqrt{k_F^2 - \left( \frac{n_x\pi}{L_x} \right)^2} \right \rfloor}_{N_y(n_x)} + 1\right ) \cos{\left( \frac{2 \pi n_x}{L_x} x \right )},
    \end{split}
\end{equation}
where $n_0^{\rm(2D)}$ is the 2D bulk density. The quantity $2N_y(n_x) + 1$ represents the number of states within a sub-band defined by a certain $n_x$, in other words, along a chord of the Fermi disk, defined by a certain discretised momentum $k_x$.

We perform the same derivation for the three-dimensional (3D) electron gas (confined in one direction, in $[0,L_x]$ and infinitely extending in the other two, simulated by periodic boundary conditions, with period $L_y$ and $L_z$). As before, standard separation of variables leads to
\begin{equation}
    \label{eq:pedagdim_3D}
    \begin{split}
    n_0(x,y,z) = & \frac{2}{L_x L_y L_z} \sum_{n_x} \sum_{n_y} \sum_{n_z}1 - \\
                 & \frac{2}{L_x L_y L_z}\sum_{n_x} \sum_{n_y} \sum_{n_z} \cos{\left( \frac{2 \pi n_x}{L_x} x \right )},
    \end{split}
\end{equation}
where $n_x$ is a natural number and $n_y$, $n_z$ are integers, representing the discretisations of the momenta, that is, $k_x = \pi n_x/L_x$, $k_y=2\pi n_y/L_y$, and $k_z=2\pi n_z/L_z$. The Fermi surface in this system is a sphere $k_x^2+k_y^2+k_z^2=k_F^2$ and all allowed states exist within the solid sphere $k_x^2+k_y^2+k_z^2 \leq k_F^2$.

In this case, we find ourselves unable to reduce the triple summation down to a single one. However we can write
\begin{equation}
    \label{eq:pedagdim_3D_2sum}
    \begin{split}
    &n_0(x,y,z) = n_0^{\rm(3D)} - \\ &
    \frac{2}{L_x L_y L_z}\sum_{n_x=1}^{N_x} \sum_{n_y=-N_y( n_x)}^{N_y( n_x)} (2 N_z(n_x,n_y) + 1) \cos{\left( \frac{2 \pi n_x}{L_x} x \right )}.
    \end{split}
\end{equation}
Above, $n_0^{\rm(3D)}$ is the 3D bulk density. The confinement in the $x$ direction discretises the Fermi solid sphere in $N_x$ disks of radius $\sqrt{k_F^2 - k_x^2}$, one for each discrete momentum. In each disk, $2N_y(n_x) + 1$ is the number of allowed sub-bands (defined by the discretised momentum $k_x$) and $2N_z(n_x,n_y) + 1$ represents the number of states along such sub-band (chord on the disk), for the particular discretised momenta $k_x$ and $k_y$.  $N_z(n_x,n_y)$ is given by
\begin{equation}
    \label{eq:pedagdim_Nz}
    N_z(n_x,n_y) = \left \lfloor \frac{L_z}{2 \pi} \sqrt{k_F^2 - \left( \frac{\pi n_x}{L_x} \right)^2 - \left( \frac{2 \pi n_y}{L_y} \right)^2 } \right \rfloor,
\end{equation}
while $N_y(n_x)$ is given as before by Eq.~(\ref{eq:pedagdim_2D_1sum}).

\begin{figure*}[ht]
    \centering
    \includegraphics[width=\linewidth]{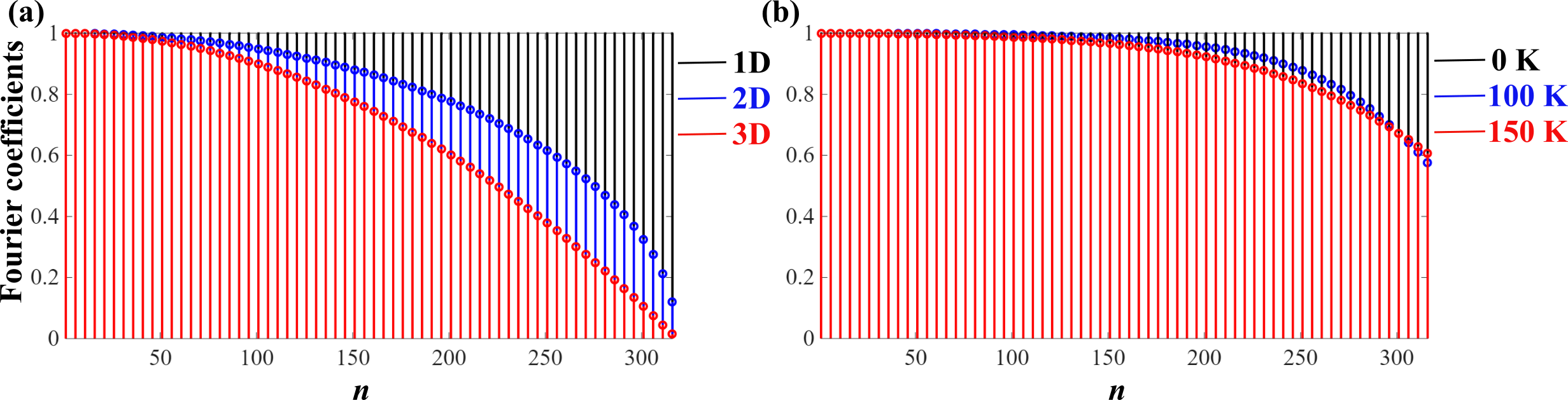}
    \caption{Strength of Fourier coefficients involved in the second terms of Eqs.~(\ref{eq:notgibbs_density}), (\ref{eq:pedagdim_2D_1sum}), (\ref{eq:pedagdim_3D_2sum}), and (\ref{eq:pedag_temp_dens}). (a) Influence of increasing dimensions on the Fourier coefficients. To facilitate comparison, the coefficients are normalised to their maximal value. We clarify that increasing dimensions leads to different amplitude also for low-$n$ modes, however the trend is as visualised here. For this simulation, $L_x=L_y=L_z=1000$ nm and $k_F = 1$\,nm$^{-1}$. (b) Influence of increasing temperature. Here the coefficients are shown unormalised. We note that increasing the temperature makes accessible additional states; we visualise here only up until $n=N$, the state corresponding to the Fermi level (at $T=0$ K). For the simulations here, $L_x=1000$ nm, $k_F = 1$\,nm$^{-1}$, while the additional states according to the criterion set in footnote 61 are 170 and 240 for $T=100$\,K and $T = 150$\,K, respectively.}
    \label{fig:coeffs}
\end{figure*}

Irrespective of the complexity that increased dimensions bring to our analysis, it is clear that Eqs.~(\ref{eq:pedagdim_2D_1sum}) and (\ref{eq:pedagdim_3D_2sum}) remain truncated cosine Fourier series, not unlike the elementary Eq.~(\ref{eq:notgibbs_density}). It is interesting then to investigate the influence of dimensions from a different lens, namely through their influence on the Fourier coefficients of each series. For the 2D and 3D case these can be considered as modifications of the 1D one~\footnote{Notice that we suppressed the $2/L$ terms appearing as multiplicative factors in all dimensions for simplicity.}, that is,
\begin{equation}
    \label{eq:pedagog_dim_Fourier_coeff}
    \begin{split}
        & a_{n_x} ^{\rm (1D)} = 1, \\ 
        & a_{n_x} ^{\rm (2D)} = \left[ 2 N_y( n_x) + 1\right ] a_{n_x} ^{\rm (1D)}, \\ 
        & a_{n_x}^{\rm (3D)} = \sum_{n_y=-N_y( n_x)}^{N_y( n_x)} \left[2 N_z(n_x,n_y) + 1\right] a_{n_x} ^{\rm (1D)},
    \end{split}
\end{equation}
with $n_x = 1, 2,\dots, N_x$. We plot them (normalised to their maximal value, which can be easily inferred that occurs when $n_x =1$) in Fig.~\ref{fig:coeffs}. In the 1D case the coefficients do not decay at all and underline the oscillatory nature of the Dirichlet kernel of Eq.~(\ref{eq:notgibbs_density}). Its oscillatory nature appears in Friedel oscillations, though not with the same mechanism as in the Gibbs phenomenon, as we discussed earlier. On the other hand, the Fourier coefficients of the higher-dimension cases demonstrate a smoothly decaying envelope. This envelope reflects the shape of constant-$k_x$ cross sections of the Fermi sea (which terminates sharply irrespective of the dimension of the electron gas). As such, for the the 1D case, we note a pure discontinuity at $|k_x|=k_F$ (reflecting the sharpness of the Fermi--Dirac distribution), which degenerates to continuity but nondifferentiability for the 2D case (the first derivative diverges). For the 3D case, the expression is sufficiently complicated to discourage further treatment. However, we can acquire intuition by using the thermodynamic limit. That is, as long as $\Delta k_y = 2\pi/L_y \to 0$ and $\Delta k_z = 2 \pi/L_z \to 0$, the inner double summation in Eq.~(\ref{eq:pedagdim_3D}) merely counts the area of a disk with radius $\sqrt{k_F^2 - k_x^2}$. This is $\pi (k_F^2 -k_x^2)$. Even though the function appears continuous with continuous derivatives, the abrupt termination at $|k_x|=k_F$ leads to a discontinuity in the first derivative. 

From the perspective and vocabulary of the Gibbs phenomenon, the faster decay rate of the Fourier coefficients masks the abrupt termination of the Fourier space (which exists always and hence induces oscillations in all cases), which is equivalent to the weaker singularity appearing at the Fermi edge. We could further loosely (and carefully) interpret the weaker and less persistent oscillations as the spatial dimensions rise as a case of application of a \emph{space filter}~\cite{Gottlieb1997}. This is a low-pass filter that  smoothly weakens the high spatial frequency modes, thus negating the effects of the sharp truncation. The geometric perspective discussed above does exactly this. In both cases, the fast oscillations of the Dirichlet kernel are suppressed depending on the efficiency of the space filter/the dimensionality. Though we find that such a parallel makes straightforward the understanding of the influence of dimensionality on Friedel oscillations without having to resort to more complicated arguments about the geometry of the Fermi space, we caution that such comparison stands exclusively as a mental aide: the modification of the  Fourier coefficients in Eq.~(\ref{eq:pedagog_dim_Fourier_coeff}) for the 2D and 3D case do not satisfy the conditions of a space filter~\cite{Gottlieb1997}; for example, they do not also affect the low-frequency modes (the amplitude changes).

\subsection{Influence of temperature}
As for dimensions, the increase of temperature is known to affect Friedel oscillations~\cite{Grassme1993,Avouris1994}. Thermal fluctuations reduce phase coherence and hence weaken both the amplitude and the range of the oscillations~\cite{Stephanovic2026}. In particular, the oscillations acquire an exponentially decaying envelope that depends on the thermal length scale $l_T \sim  \hbar\upsilon_F/k_B T$, where $T$ is the temperature, $k_B$ is the Boltzmann constant, $\hbar$ is the reduced Planck constant, and $\upsilon_F$ is the Fermi velocity~\cite{Grassme1993}.

Equivalently, the phenomenon stems from the smoothing of the Fermi edge~\cite{Grassme1993}. This can be seen from the shape of the Fermi--Dirac distribution, which is
\begin{equation}
    \label{eq:pedag_temp_fermi}
    f_k = \frac{1}{1 + \exp{[(E_k - \mu)/k_B T]}},
\end{equation}
where $\mu$ is the chemical potential and $E_k$ are the discretised energy levels of the system. We can readily repeat all the derivations pertaining  to the 1D electron gas of Sec.~\ref{sec:gibbsfriedel}, yielding
\begin{equation}
    \label{eq:pedag_temp_dens}
    \begin{split}
    & n_0(x) = \frac{2}{L} \sum_{n=1}^{N'}\frac{1}{1 + \exp{[(E_n - \mu)/k_B T]}} - \\
    & \frac{2}{L} \sum_{n=1}^{N'} \frac{1}{1 + \exp{[(E_k - \mu)/k_B T]}}\cos{\left( \frac{2 \pi n}{L} x \right )}.
    \end{split}
\end{equation}
Above, the discretised energy levels are given by $E_n= \hbar^2 \pi^2 n^2/(2m_e L^2)$, where $m_e$ is the electron mass, while the chemical potential is calculated in such a manner that the total number of electrons in the system is conserved~\footnote{ \label{fn:chem}Some details on its calculation are in order. We solve iteratively the equation $N = \sum_{n=1}^{N'} f_n$, using the approximation $\mu = E_F$ as a reasonable first guess. We perform a convergence study per temperature point to investigate which $N' > N$ ($N'$ is the number of states including those thermally excited beyond the Fermi level) is sufficient for the chemical potential to converge, in the sense that the relative error in the chemical potential between consecutive values of $N'$ is less than $1 \%$.}~\cite{Kittel2005}. The expression above simplifies to that of Eq.~(\ref{eq:notgibbs_density}) when $k_B T \to 0$. The upper limit of the summation is a critical quantity. The sharp spectral cutoff is here replaced by a weighting procedure, states beyond the Fermi level are excited, however are of increasingly low probability. In that sense, $N'$ is the lowest number of states that captures all sufficiently strong contributions to the density (see footnote 62 for more details). We have once again described an operation that is similar to applying filtering to the nondecaying Fourier coefficients of the absolute-zero 1D case; the sigmoid profile of the Fermi--Dirac distribution guarantees that lower-frequency modes will not be appreciably affected, however high-frequency ones will be suppressed, explaining directly the phenomenology. What we observe is still the relaxation of the Dirichlet kernel. We further note that our analogy to filtering is quite stronger here, as it is clear that the new Fourier coefficients $a_n^{(T \neq 0)} = f_n a_n^{(T=0)}$ are such that $a_n^{(T \neq 0)} \simeq a_n^{(T=0)}$, for $n \ll N'$, while  $a_n^{(T \neq 0)} < a_n^{(T=0)}$ otherwise (though we clarify that the Fermi--Dirac distribution does not satisfy the rigorous definition of a filter~\cite{Gottlieb1997}, while increase in temperature leads to an increase of the terms added in Eq.~(\ref{eq:pedag_temp_dens})). We plot the strength of the Fourier coefficients for three increasing temperatures in Fig.~\ref{fig:coeffs}(b).

\section{\label{sec:concl}Conclusion}
In this work, we interrogated the relationship between the Gibbs phenomenon and Friedel oscillations, inspired by works that have already pointed out similarities. Even though we found that these are very noticeable, and echo truncations of the Fourier/Fermi space, which is a necessary condition for both, Friedel oscillations cannot be called Gibbs phenomenon. The main reasons are the inability to find a final, limiting function (for the unnormalised density) and the dependence of the coefficients on the number of terms used in the series (for the normalised density). Nonetheless, we find that the parallels are sufficient to allow for swift (even if not very rigorous) interpretations of behaviours of Friedel oscillations. For both the cases of increased dimensionality of the electron gas and increased temperature, the resulting weaker oscillations are interpreted as the result of accelerating the decay rate of Fourier coefficients,  a behaviour reminiscent of a standard attempt to alleviate the Gibbs phenomenon, namely space filters. We hope that aside from clarifying an ongoing discussion, the current work offers an entry-level explanation of salient behaviours of Friedel oscillations, to an audience that is unaccustomed to using their vocabulary.

\begin{acknowledgments}
We acknowledge support from the Independent Research Fund Denmark (Grant No.~5281-00155B). The Center for Polariton-driven Light-Matter Interactions (POLIMA) is funded by the Danish National Research Foundation (Project No. DNRF165). We are thankful to Line Jelver, P. Andr\'{e} D. Gon\c{c}alves, and N. Asger Mortensen for helpful discussions.
\end{acknowledgments}

\bibliography{apssamp}% Produces the bibliography via BibTeX.

\end{document}